\documentclass[runningheads,a4paper]{llncs}
\authorrunning{Gawlik, Koppe, Kollenda, Pawlowski, Garmany and Holz}
\usepackage{siunitx}
\usepackage{caption}
\usepackage{booktabs}
\usepackage[english]{babel}
\usepackage{subfig}
\usepackage{cite}
\usepackage{url}
\usepackage{listings}
\usepackage[plain]{algorithm}
\usepackage{algorithmic}
\usepackage{amssymb}
\usepackage{pifont}
\usepackage{rotating}
\usepackage{multirow,tabularx}
\usepackage{flushend}
\usepackage{fp}
\usepackage{blindtext}
\usepackage{mathrsfs}
\usepackage{empheq}
\usepackage{paralist}

\usepackage{relsize}
\usepackage{lipsum}
\usepackage{balance}
\usepackage{placeins}

\usepackage{graphicx} 
\usepackage[hidelinks]{hyperref}
\usepackage[usenames,dvipsnames]{color}
\usepackage{amsmath}
\usepackage{lscape}
\usepackage{multicol}

\usepackage{moresize}

\usepackage{anyfontsize}

\usepackage{marginnote}

\newcommand{\comment}[1]{}

\newcommand{\toolname}{\textsc{Detile}\xspace} %

\newcommand \infoleak {information leak\xspace}
\newcommand \raslr {per process re-randomization\xspace}

\newcommand \orig {master\xspace}
\newcommand \twin {twin\xspace}
\newcommand \dexec {dual process execution\xspace}

\newcommand{\mytitle}{\toolname{}: Fine-Grained Information Leak Detection \\ in Script Engines}

\institute{Horst G\"ortz Institute for IT-Security (HGI)\\ Ruhr-Universit\"at Bochum, Germany\\\email{\{firstname.lastname\}@rub.de}} %
    \title{\mytitle}
    
\begin{document}

\author{Robert Gawlik, Philipp Koppe, Benjamin Kollenda, \\ Andre Pawlowski, Behrad Garmany \and Thorsten Holz}

\date{}

\maketitle

\begin{abstract}
\emph{Memory disclosure attacks} play an important role in the 
exploitation of memory corruption vulnerabilities. By analyzing recent
research, we observe that bypasses of defensive solutions that enforce control-flow
integrity or attempt to detect return-oriented programming require memory
disclosure attacks as a fundamental first step.
However, research lags behind in detecting such information leaks.

In this paper, we tackle this problem and present a system for fine-grained,
automated detection of memory disclosure attacks against scripting engines.
The basic insight is as follows: scripting languages, such as
JavaScript in web browsers, are strictly sandboxed. They must not provide any
insights about the memory layout in their contexts.  In fact, \emph{any} such
information potentially represents an ongoing memory disclosure attack.  Hence,
to detect information leaks, our system creates a clone of the scripting
engine process with a re-randomized memory layout. The clone is instrumented to
be synchronized with the original process.  Any inconsistency in the script
contexts of both processes appears when a memory disclosure was conducted to
leak information about the memory layout. Based on this detection approach, we
have designed and implemented \toolname{} (\underline{det}ection of \underline{i}nformation \underline{le}aks), a prototype for the JavaScript engine in
Microsoft's Internet Explorer 10/11 on Windows 8.0/8.1. An empirical
evaluation shows that our tool can successfully detect memory disclosure
attacks even against this proprietary software.
\end{abstract}

\section{Introduction}

Over the last years, many different techniques were developed to prevent attacks that exploit spatial and temporal memory corruption vulnerabilities (see for example the survey by Szekeres et al.~\cite{Szekeres:2013:EWM}). As a result, modern operating systems deploy a wide range of defense methods to impede a successful attack.
For example, \emph{Data Execution Prevention} (DEP)~\cite{molnar2003exec} marks data as
non-executable and thus an attacker is prohibited from injecting data into a
vulnerable application that is later on interpreted as code.
Furthermore, \emph{Address Space Layout Randomization} (ASLR)~\cite{russinovich2005microsoft}
randomizes the memory layout
either once during the boot process or every time a process is started.
Since the attacker lacks information about the exact memory layout, it is harder for
her to predict where her shellcode or reusable code are located.

Besides these widely deployed techniques, many other defenses were proposed in the literature in the last years~\cite{Szekeres:2013:EWM}. Most notably, the enforcement of \emph{control flow integrity} (CFI) is a promising technique to prevent a whole class of memory corruption vulnerabilities~\cite{abadi2005control}. The basic idea behind CFI is to verify that each control flow transfer leads to a valid target based on a control flow graph that is either statically pre-computed or dynamically generated. Several implementations of CFI with different design constraints, security goals, and performance overheads were published (e.g.,~\cite{Erlingsson:2006:XFI, zhang2013practical, zhang2013bincfi}).

A general observation is that the first step in modern attacks is based on a \emph{memory disclosure attack} (also referred to as \emph{information leak}): the adversary finds a way to read a (raw) memory pointer to learn some information about the virtual address space of the vulnerable program. Generally speaking, the attacker can then de-randomize the address space based on this leaked pointer (thus bypassing ASLR),
use ROP to bypass DEP,
and finally execute shellcode of her choice. Modern exploits leverage information leaks as a fundamental primitive. Furthermore, recent CFI and ROP defense bypasses use memory disclosures as well. For example, Snow et al. introduced \emph{Just-In-Time Code Reuse} attacks (JIT-ROP~\cite{snow2013just}) to bypass fine-grained ASLR implementations by repeatedly utilizing an information leak. \emph{G-Free}~\cite{onarlioglu2010g}, a compiler-based approach against any ROP attack, was recently circumvented by Athanasakis et al.~\cite{athanasakis2015devil}. Their technique requires successive information leaks to disclose enough needed information. G\"{o}kta\c{s} et al. demonstrated several bypasses of proposed ROP defenses and their exploit needs an information leak as a first step~\cite{goktacs2014size}. An information leak is also needed by Song et al., who showed that dynamic code generation is vulnerable to code injection attacks~\cite{song2015exploiting}. Similarily, \emph{Counterfeit Object-oriented Programming} (COOP~\cite{schuster2015counterfeit}) needs to disclose the location of \emph{vtables} to mount a subsequent control-flow hijacking attack by reusing them. Disclosures are also utilized by \emph{memory oracles} to weaken various defenses~\cite{gawlik2016enabling}. \emph{All} of these offensive bypasses utilized an information leak as a first step and implemented the attack against a web browser.

\begin{table*}[t!]
    \centering
    \caption{
        Defenses and offensive approaches
        utilizing an information leak in browsers to weaken or bypass the
        specific defense. All mentioned attacks are mitigated by \toolname{}.
    }
    ~\\

      \tiny
    \begin{tabular}{ l c c c}

			\toprule

			\textbf{Protection flavor} & \textbf{Defense} & \textbf{Weakened/Bypassed by} & \textbf{Mitigated by} \toolname{} \\

			\toprule

			Address randomization & Fine-grained ASLR~\cite{hiser2012ilr} & Just-In-Time Code Reuse~\cite{snow2013just} & $\surd$\\

			\midrule

								  & RopGuard~\cite{Frantic_ROP}, & Size Does Matter~\cite{goktacs2014size}, & \\
			Code-reuse protection & KBouncer~\cite{Pappas:2013}, & Anti-ROP Evaluation~\cite{schuster2014evaluating}, & $\surd$ \\
								  & ROPecker~\cite{cheng:ropecker:2014} & COOP~\cite{schuster2015counterfeit} & \\

			\midrule

			Code-reuse protection & G-Free~\cite{onarlioglu2010g} & Browser JIT Defense Bypass~\cite{athanasakis2015devil}, & $\surd$\\
			 & & COOP~\cite{schuster2015counterfeit} & \\

			\midrule

			Coarse-grained CFI & CCFIR~\cite{zhang2013practical}, & Stitching the Gadgets~\cite{davi2014stitching}, & \\
							   & BinCFI~\cite{zhang2013bincfi} & Out of Control~\cite{goktas2014out}, & $\surd$ \\
							   & & COOP~\cite{schuster2015counterfeit} & \\

			\midrule

			Fine-grained CFI & IFCC~\cite{tice2014enforcing}, & Losing Control~\cite{conti2015losing} & $\surd$\\
							 & VTV~\cite{tice2014enforcing} & & \\

			\midrule

							   & & Vtable disclosure~\cite{davi2015isomeron}, & \\
			Information-hiding & Oxymoron~\cite{backes2014oxymoron} & Crash-Resistance~\cite{gawlik2016enabling} & $\surd$\\
							   &                                    & COOP~\cite{schuster2015counterfeit} & \\

			\midrule

			Information-hiding & CPI linear region~\cite{kuznetsov2014code} & Crash-Resistance~\cite{gawlik2016enabling} & $\surd$\\

			\midrule

			Execution randomization & Isomeron~\cite{davi2015isomeron} & Crash-Resistance~\cite{gawlik2016enabling} & $\surd$\\

			\midrule

			Randomization/Information-hiding & Readactor~\cite{crane2015readactor} & Crash-Resistance~\cite{gawlik2016enabling}, & $\surd$\\
								& & COOP~\cite{schuster2015counterfeit} & \\

			\bottomrule

    \end{tabular}
    \label{tab:cat_and_mouse}
\end{table*}
Another general observation is that script engines in web browsers are commonly utilized by adversaries to abuse information leaks in practice.
Browser vulnerabilities are prevalent and as the yearly \emph{pwn2own} competition shows, researchers successfully use them to take control of the machine. Notably, most of these attacks are based on vulnerabilities that create an information leak utilizing the script engine.

In this paper, we take these observations into account and propose a technique for fine-grained, automated detection of memory disclosure attacks against script engines at runtime. Our approach is based on the insight that information leaks are leveraged by state-of-the-art exploits to learn the placement of modules---and thereby code sections---in the virtual address space in order to bypass ASLR.
Any sandboxed script context is forbidden to contain memory information, i.e., no script variable is allowed to provide a memory pointer.
As such, a viable approach to detect information leaks is to create a clone of the to be protected process with a re-randomized address space layout, which is instrumented to be synchronized with the original process. An inconsistency in the script contexts of both processes can only occur when a memory disclosure vulnerability was exploited to gain information about the memory layout. In such a case, the two processes can be halted to prevent further execution of the malicious script. An overview of bypassed defenses by specific attacks which are mitigated by our approach is shown in Table~\ref{tab:cat_and_mouse}.

We have implemented a prototype of our technique in a tool called \toolname{} (\underline{det}ection of \underline{i}nformation \underline{le}aks). We extended Internet Explorer 10/11 (IE) on Windows~8.0/8.1 to create a synchronized clone of each tab and enforce the information leak checks. We chose this software mainly due to two reasons. First, IE is an attractive target for attackers as the large number of vulnerabilities indicates. Second, IE and Windows pose several interesting technical challenges since it is a proprietary binary system that we need to instrument and it lacks fine-grained ASLR.
Evaluation results show that our prototype is able to re-randomize single processes without significant computational impact. Additionally, running IE with our re-randomization and information leak detection engine imposes a performance hit of $\sim$17\% on average.
Furthermore, empirical tests with real-world exploits also indicate that our approach is usable to unravel modern and unknown exploits which target browsers and utilize memory disclosures.

\smallskip
\noindent
In summary, our main contributions in this paper are:
\begin{itemize}
    \item We present a system to tackle the problem of {\infoleak}s, which are frequently used in practice by attackers as an exploit primitive. More specifically, we propose a concept for fine-grained, automated detection of {\infoleak}s with \raslr, \dexec, and process synchronization. An extended version of this paper with more technical details is available as a technical report~\cite{TR_INFO_LEAKS}.
    \item We show that dual execution of highly complex, binary-only software such as Microsoft's Internet Explorer is possible without access to the source code, whereby two executing instances operate deterministic to each other.
    \item We implemented a prototype for IE 10/11 on Windows 8.0/8.1.
      We show that our tool can successfully detect several real-world exploits, while producing no alerts on highly complex, real-world websites.
\end{itemize}

\section{Technical Background}

In the following, we briefly introduce several concepts needed to understand the challenges we were confronted with when developing \toolname{}.

\subsection{N-Variant Systems}
\label{discussion:nvariant}
\emph{N-Variant} or \emph{Multi-Execution} systems evolved from
fault-tolerant environments to mitigation systems against
security critical vulnerabilities~\cite{Cox:2006:NSS, bruschi2007diversified,
volckaert2015cloning, varan:asplos15}. Our concept of \toolname{}
incorporates similar ideas like dual process execution and dual process
synchronization. However, our approach is constructed specifically for
scripting engines, and thus, is more fine-grained: While \toolname{} operates
and synchronizes processes on the scripting interpreter's bytecode level,
n-variant systems intercept only at the system call level. One drawback for
these conventional systems is that they are prone to \emph{Just-In-Time
Code-Reuse} (JIT-ROP~\cite{snow2013just}) and \emph{Counterfeit Object-oriented
Programming} (COOP~\cite{schuster2015counterfeit}) attacks, while \toolname{} is able to detect
these (see Sections~\ref{overview:concept} and~\ref{_sec:related_n_variant} and for details).

\subsection{Windows ASLR Internals}
\label{background:aslr}
\emph{Address Space Layout Randomization} (ASLR) is a well-known security mechanism that involves the  randomization of stacks, heaps, and loaded images in the virtual address space. Its purpose is to leave an attacker with no knowledge about the virtual memory space in which code and data lives. Combined with DEP, ASLR makes remote system exploitation through memory corruption techniques a much harder task. While brute-force attacks against services that automatically restart are possible~\cite{bittau2014hacking}, such attacks are typically not viable in practice against web browsers.

In Windows, whenever an image is loaded into the virtual address space, a
section object is created, which represents a section of memory. These objects
are managed system-wide and can be shared among all processes. Once a DLL is
loaded, its section object remains permanent as long as processes are
referencing it. This concept has the benefit that relocation takes place once
and whenever a process needs to load a DLL, its section object is reused and the
view of the section is mapped into the virtual address space of the process,
making the memory section visible. This way, physical memory is shared among all
processes that load a specific DLL whose section object is already present. In
particular, as long as the virtual address is not occupied, each image is loaded
at the same virtual address among all running usermode processes.
\subsection{WOW64 Subsystem Overview}
\label{background:wow64}

64-bit operating systems are the systems of choice for today's users: 64-bit
processors are widely used in practice, and hence Microsoft Windows 7 and later versions are usually running in the 64-bit
version on typical desktop systems. However, most third-party
applications are distributed in their 32-bit form. This is for example the case for
Mozilla Firefox, and \comment{surprisingly }also for parts of
Microsoft's Internet Explorer. As our
framework should protect against widely attacked targets, it needs to support
32-bit and 64-bit processes. Therefore, the \emph{Windows On Windows 64} (shortened as \emph{WOW64})
emulation layer plays an important role, as it allows legacy 32-bit applications
to run on modern 64-bit Windows systems.

Executing a user-mode 32-bit application instructs the kernel to create a WOW64
process. According to our observations, it creates the program's address space and maps
the 64-bit and 32-bit \emph{NT Layer DLL} (\texttt{ntdll.dll}) and the main
executable into it. Even when a program may have been started in suspended mode, these three
modules are already available. Afterwards, WOW64 layer DLLs are mapped,
which mediate several necessary transitions between 64-bit and 32-bit at
runtime~\cite{russinovich2005microsoft}.
Subsequent 32-bit DLLs are mapped into the address space via \texttt{LdrLoadDll} of the
32-bit \texttt{ntdll.dll}. The first of them is \texttt{kernel32.dll}.
The loader assures that it is mapped to the same address in each WOW64 process
system wide, using a unique address per reboot. It therefore compares its name to
the hardcoded ``KERNEL32.DLL'' string in \texttt{ntdll.dll} upon
loading. If the loader is not able to map it to its
preferred base address, process initialization fails with a conflicting address
error. As process based re-randomization plays a crucial role in our framework,
this issue is handled such that each process contains its \texttt{kernel32.dll}
at a different base address (see Section~\ref{implementation:rerand}).
After mapping \texttt{kernel32.dll}, all other needed 32-bit DLLs are mapped into the address space.

\subsection{Internet Explorer Architecture}
\label{background:ie}
IE is developed as multi-process application~\cite{zeigler2008lcie}.
That means, a 64-bit main frame process governs several 32-bit WOW64 tab processes, which are
isolated from each other. The frame process runs with a medium integrity level
and isolated tab processes run with low integrity levels. Hence, tab processes
are restricted and forbidden to access all resources of processes with higher
integrity levels~\cite{ms2014integrity}.
This architecture implies that websites opened in new tabs can
lead to the start of new tab processes. These have to incorporate our protection
in order to protect IE as complete application against \infoleak{}s (see
Section~\ref{implementation}).

\label{background:iejs}
\subsection{Scripting Engines}
In the context of IE, mainly two scripting engines are relevant and we briefly introduce both.
\paragraph{Internet Explorer Chakra.}
With the release of Internet Explorer 9, a new JavaScript engine called
\emph{Chakra} was introduced. Since Internet Explorer 11, Chakra exports a
documented API which enables developers to embed the engine into their own
applications. However, IE still uses the undocumented internal COM interface.
Nevertheless, some Chakra internals were learned from the official API.
The engine supports just-in-time (JIT) compiling of JavaScript bytecode
to speed up execution. Typed arrays like integer arrays are stored as
native arrays in heap memory along with metadata to accelerate element
access. Script code is translated to JS bytecode on demand in a function-wise
manner to minimize memory footprint and avoid generating unused bytecode.
The bytecode is interpreted within a loop, whereby undocumented \emph{opcodes}
govern the execution of native functions within a switch statement. Dependent on
the opcode, the desired JavaScript functionality is achieved with native
code.
\paragraph{ActionScript Virtual Machine (AVM).}
The \emph{Adobe Flash} plugin
for browsers and especially for IE is a widely attacked target. Scripts written in
\emph{ActionScript} are interpreted or JIT-compiled to native code by the AVM.
There is much unofficial documentation about its internals~\cite{blazakis2010interpreter,insideavm2012}. Most
importantly, it is possible to intercept \emph{each} ActionScript method with
available tools~\cite{avmfunprofit2014}. Thus, no matter whether bytecode is interpreted by the
opcode handlers or JIT code is executed, we are able to instrument the AVM.

\subsection{Adversarial Capabilities}
\label{background:attacker}

Memory disclosure attacks are an increasingly used technique for the exploitation of software
vulnerabilities~\cite{snow2013just, serna2012info, strackx2009breaking}. In the
presence of full ASLR, DEP, CFI, or ROP defenses, the attacker has no anchor to a memory address to
jump to, even if in control of the instruction pointer. This is the moment where
\infoleak{}s come into play: an attacker needs to read---in any way possible---a
raw memory pointer in order to gain a foothold into the native virtual address
space of the vulnerable program.
As soon as the attacker can read process
memory, she can learn the base addresses of
loaded modules. Then, any code reuse
primitives can be conducted to exploit a vulnerability in order to bypass DEP,
ASLR, CFI~\cite{davi2014stitching} and ROP defenses~\cite{carlini2014rop, goktacs2014size}. Another possibility is
to leak code directly in order to initiate an attack and bypass ASLR~\cite{snow2013just}.
Other mitigations like Microsoft's Enhanced Mitigation Experience
Toolkit (EMET)~\cite{ms2014emet} cannot withstand capabilities of sophisticated attackers.

For applications with scripting capabilities, untrusted contexts are sandboxed
(e.g., JavaScript in web browsers) and must not provide memory information.
Thus, attackers use different vulnerabilities
to leak memory information into that context~\cite{goktas2014out,yan2014art,
serna2012info}. We assume that the program we want to protect suffers from such
a memory corruption vulnerability that allows the adversary to corrupt memory
objects. In fact, a study shows that \emph{any} type of memory error can be transformed into an
\infoleak{}~\cite{Szekeres:2013:EWM}. Furthermore, we assume that the
attacker uses a scripting environment to leverage the obtained memory
disclosure information at runtime for her malicious computations. This is consistent with modern exploits in
academic research~\cite{goktas2014out, davi2014stitching, goktacs2014size,
carlini2014rop, schuster2014evaluating} as well as in-the-wild~\cite{yu2014write,
yu2014rops, yan2014art, vupenFox, browserweakestbyte}.
Our goal is to protect script engines against such powerful, yet realistic adversaries.
\section{System Overview}

In the following, we explain our approach to tackle the challenge of detecting information leaks in script engines.
Hence, we introduce the needed building blocks, namely \raslr and \dexec. %

\subsection{Main Concept}
\label{overview:concept}
As described above, {\infoleak}s manifest themselves in the
form of memory information inside a context which
must not reveal such insights.
In our case, this is any script context inside an application:
high-level variables and content in a script
must not contain memory pointers, which attackers could use to deduce
image base addresses of loaded modules.

Unfortunately, a legitimate number and a memory pointer in data bytes received via a scripting function are indistinguishable.
This leads us to the following assumption: a memory disclosure attack yields a
memory pointer, which may be surrounded by legitimate data. The same targeted
memory disclosure, when applied to a differently \emph{randomized}, but
otherwise \emph{identical} process, will yield the same legitimate data, but a
\emph{different} memory pointer. Due to the varying base addresses of modules,
different heap and stack addresses, a memory pointer will have a different
address in the second process than in the first process. Thus, a \orig process
and a cloned \twin process---with different address space layout randomization---can 
be executed synchronized side-by-side and perform identical operations,
e.g., execute a specific JavaScript function. In benign cases, the same data
getting into the script context is equal for both processes.
When comparing
the received data of one process to the same data received in the second
process, the only difference can arise because of a leaked memory pointer
pointing to equal memory, but having a \emph{different} address.
In order to compare the data of the \orig and \twin process, we have to instrument
the interpreter loop of the script engine.
We can instrument the \texttt{call}
and \texttt{return} bytecodes to precisely check all outgoing data and therefore
to detect an \infoleak{}.

\begin{figure}[!htb]
	\centering
        \includegraphics[width=0.7\linewidth]{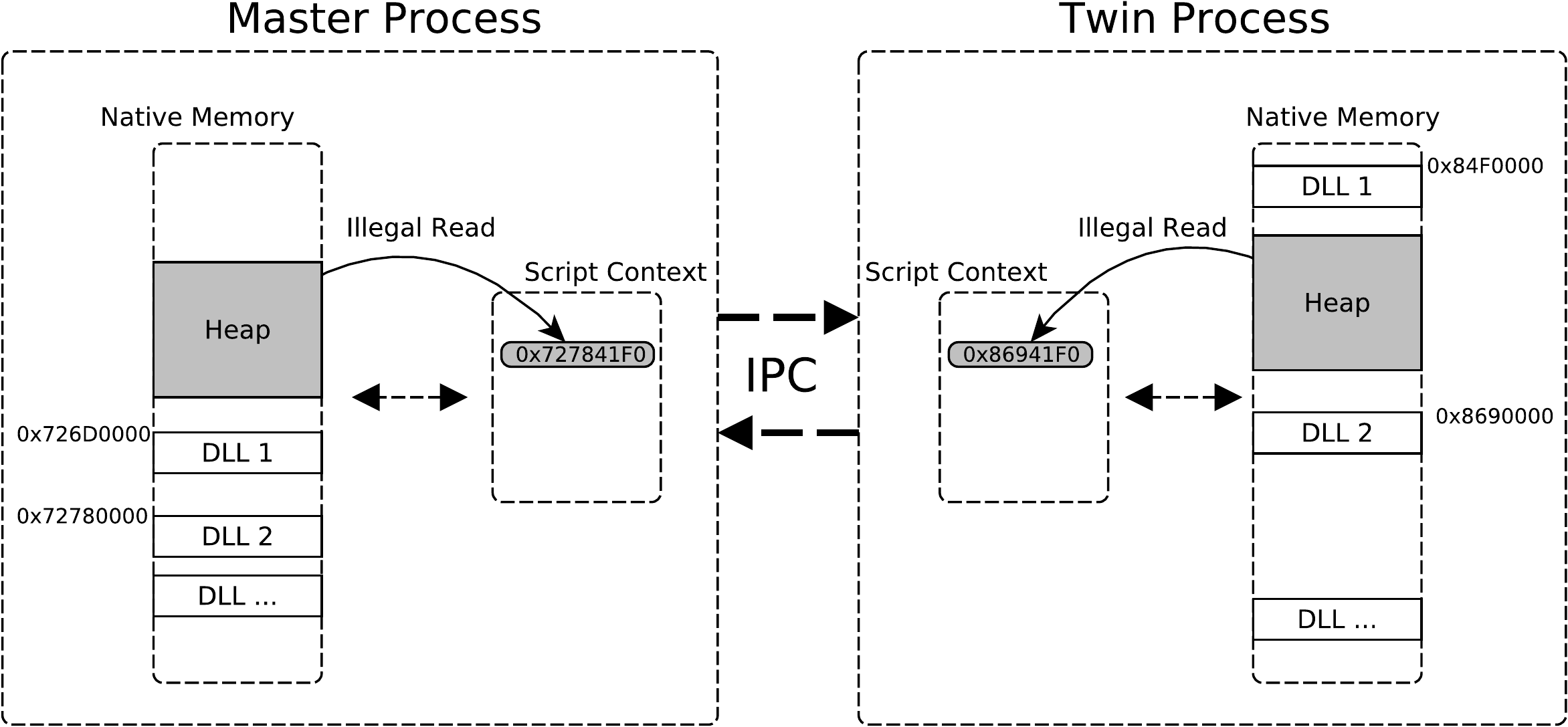}
    \caption{Overview of our main \infoleak detection concept: The \orig
        process is synchronized with a re-randomized, but otherwise identical
        \twin process. If a memory disclosure attack is conducted in the \orig
        \comment{process}, it appears as well in the \twin \comment{process}.  Due to the different
        randomization, the disclosure attack manifests itself in different
        data flowing into the script context and can be detected
        (\texttt{0x727841F0} vs. \texttt{0x86941F0})}
    \label{fig:leak_detector}
\end{figure}

Based on this principle, our prototype system launches the same
script engine process twice with diverse memory layouts (see also Figure~\ref{fig:leak_detector}). The script engines are coupled to
run in sync which enables checking for \infoleak{}s. In spirit, this is similar to n-variant systems~\cite{Cox:2006:NSS,bruschi2007diversified}
and multi-execution based approaches~\cite{devriese2010noninterference,capizzi2008preventing,croft2011towards}. However, our approach is more fine-grained since it checks and synchronizes the processed data on the bytecode level of the script context and is capable of detecting the actual information leak, instead of merely detecting an artifact of a successful compromise (i.e., divergence in the control flow).

\subsection{Per Process Re-randomization}
\label{overview:rerandomization}
To overcome the dilemma of modules having equal base addresses in different
processes, we collect all base addresses of modules a process loads during its
runtime.  We refer to this first process, which is launched, as \emph{\orig}
process. A second process instance of the application known as the \emph{\twin}
process is spawned. Upon its initialization, the base addresses gained from the
\orig are occupied in the virtual address space of the \twin. This forces the
image loader to map the modules to other addresses than in the \orig process, as
they are already allocated\comment{ (see Figure~\ref{fig:mapview_rand})}. We save us the time
and trouble to re-randomize the stack and heap process-wise, as modern operating
systems (e.g., Windows 8 on 64-bit) support it natively. Finally, we establish an
\emph{inter-process communication} (IPC) bridge between the \orig and \twin process.
This enables synchronized execution between them and comparison of data flows
into their script contexts.
\subsection{Dual Process Synchronization}
\label{overview:dualprocess}
After the re-randomization phase, both processes are ready to start
execution at their identical entrypoints. After exchanging a handshake, both
resume execution. In order to achieve comparable data for information leak
checking, the executions of script interpreters in both processes have to be
synchronized precisely.
This is accomplished by intercepting an interpreter's
native methods. Additionally, we install hooks inside the bytecode interpreter loop at
positions where opcodes are interpreted and corresponding native functions are
called.
Thus, we perceive any high-level script method call at its binary level.
The \orig drives execution and these hooks are the points where the \orig and
\twin process are synchronized via IPC. We check for
\infoleak{}s by comparing binary data which returns as high-level
data into the script context.
All input data the \orig loads are stored in a
cache and replayed to the \twin process to ensure they operate on the same
source (e.g., web pages a browser loads).
Built-in script functions that potentially introduce entropy (e.g.,
\texttt{Math.random}, \texttt{Date.now}, and \texttt{window.screenX} in JavaScript) interfere
with our deployed detection mechanism, since they generate values inside the
script context that are different from each other in the \orig and \twin
processes, respectively.  Additionally, they may induce a divergent script
control flow. Both occurrences would be falsely detected as memory disclosure.
Thus we also synchronize the entropy of both processes
by copying the generated value from the \orig to the \twin process.
This way the \twin process continues working on the same data as the \orig process and
we are creating a co-deterministic script execution.

\section{Implementation Details}
\label{implementation}
Based on the concepts of \raslr and \dexec, we implemented a tool called \toolname
for Windows 8.0 and 8.1 64-bit. The current prototype is able to re-randomize on a per process
basis and instrument Internet Explorer 10 and 11 to run in \dexec mode.

\subsection{Duplication and Re-randomization}
\label{implementation:rerand}
In order to re-randomize processes and load images at different base
addresses, we developed a duplicator which creates a program's \orig
process.  It enumerates the \orig's initial loaded images with the help of the
Windows API (\texttt{CreateToolHelp32Snapshot}) before the master starts
execution. Then, the twin process is created in suspended mode, and a page is allocated in the \twin at all 
addresses of previously gathered image bases.
We then need to trick the Windows loader into mapping \texttt{kernel32.dll} at a
different base in the \twin. This is achieved by leveraging the DebugAPI and via
manipulating parameters at calls of \texttt{RtlEqualUnicodeString} in the 32-bit
loader in the \texttt{ntdll.dll}. This way, the loader believes that a
\emph{different} DLL than \texttt{kernel32.dll} is going to be initialized and
allows the mapping to a different base. It is the first DLL which is loaded
after the WOW64 subsystem.  Thus, all subsequent libraries that are loaded and
import functions from \texttt{kernel32.dll} have no problems to resolve their
dependencies using the remapped \texttt{kernel32.dll}. The loader maps them to
different addresses, as their preferred base addresses are reserved. Although
the DebugAPI is used, all steps run in a fully automated way.
As a next step, the DebugAPI is detached and the main image is remapped to a different
address. As it is already mapped even in suspended processes, this has to be done specifically.
Additionally,
\texttt{LdrLoadDll} in the \twin process is detoured to
intercept new library loads and map incoming images to different addresses than
in the \orig.
Technical details about our remapping can be found in the technical report~\cite{TR_INFO_LEAKS}.

We were not able to re-randomize \texttt{ntdll.dll} because it is mapped into the
virtual address space very early in the process creation procedure. Attempts to
remap \texttt{ntdll.dll} later on did not succeed due to callbacks invoked by the kernel.
The implications of a non re-randomized \texttt{ntdll.dll} are discussed in Section~\ref{discussion:limit}.

Note that this design works also with pure 64-bit processes.
However, frequently attacked applications like tab processes of Internet Explorer are 32-bit
and are running in the WOW64 subsystem. Hence, our framework has to protect them as well. The following explains how \toolname{} achieves this support.

\begin{figure}[tb]
	\centering
	    \includegraphics[width=0.5\linewidth]{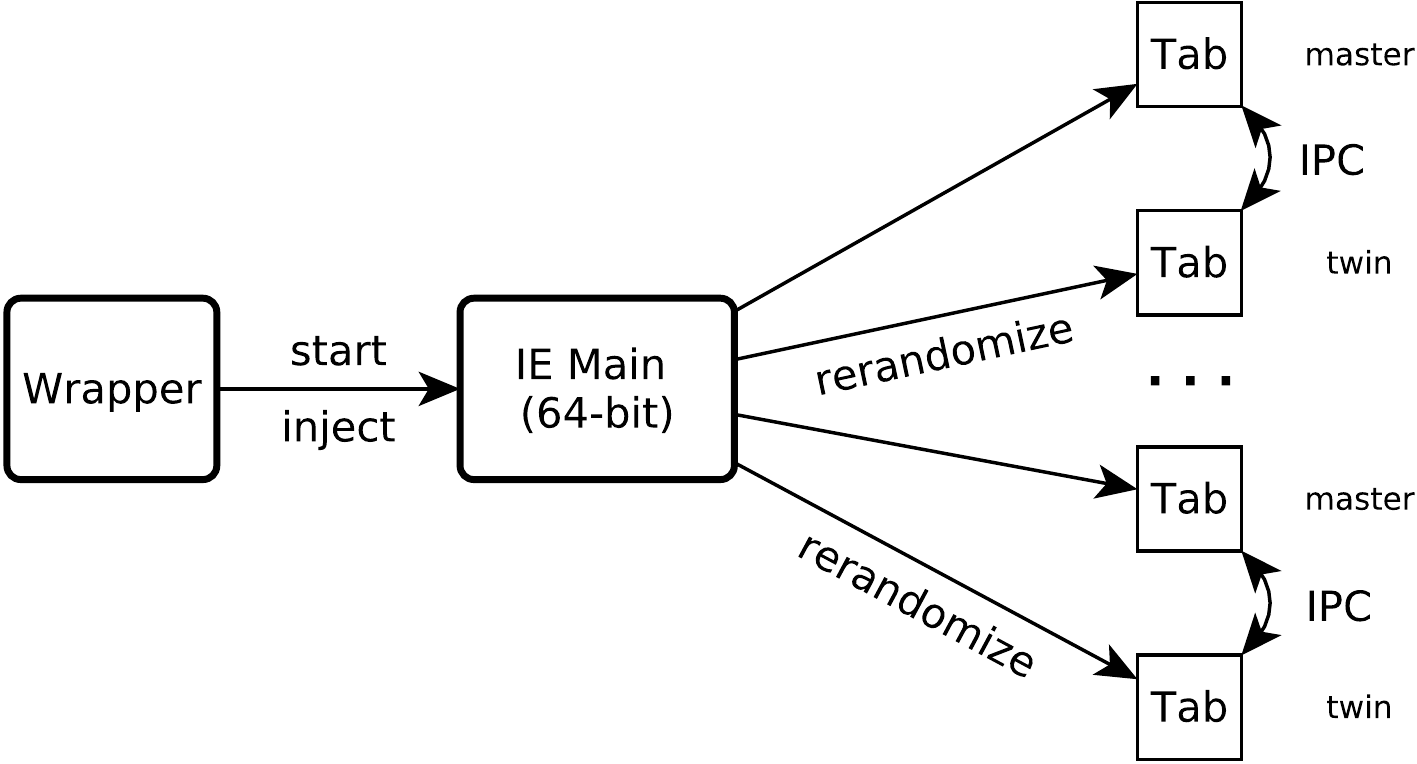}
	\caption{\toolname running with Internet Explorer. A 64-bit duplicator library is
        injected into the main IE frame process to enable it creating and
        rerandomizing \twin tab processes for each \orig tab process, by itself. The main IE
        frame also injects a 32-bit DLL into each tab process to allow
        synchronization, communication between \orig and \twin, and \infoleak detection.
    }
	\label{fig:duplicator}
\end{figure}

While the above explained logic is sufficient to duplicate and re-randomize a single-process program, additional
measures have to be taken in the case of multi-process architecture
applications like Internet Explorer. Therefore, we developed a wrapper which starts the
64-bit main IE frame process and injects a 64-bit library, which we named duplicator
library (see Figure~\ref{fig:duplicator}). 
This way, we modify the frame
process, such that each time a tab process is started by the frame process, a second tab process is
spawned. The first becomes the \orig, the second the \twin. This is achieved via
detouring and modifying the process creation of the IE frame. Additionally, our
above explained re-randomization logic is incorporated into the duplicator
library to allow the main IE frame process itself to re-randomize its spawned
\twin{}s at creation time.
To protect each new tab which is run by the IE frame, we ensure that each tab is run
in a new process and gets a \twin.
To enable communication, synchronization, and detection of \infoleak{}s, the
duplicator injects also a 32-bit library into the \orig and the \twin upon their
creation by the main IE frame process.

\subsection{Synchronization}

We designed our prototype to be contained in a DLL which is loaded into both
target instances. To reliably intercept all script execution, we hook
\texttt{LdrLoadDll} to initialize our synchronization as early as possible
once the engine has been loaded. After determining the role (\orig or
\twin), the processes exchange a short handshake and wait for events from the interpreter
instrumentation. While most of our work is focused on the scripting engine, we also
instrument parts of \texttt{wininet.dll} to provide basic proxy
functionality. The \twin receives an exact copy of the web data sent to the \orig
to ensure the same code is executed.

\paragraph{Entropy Normalization.}
The synchronization of script execution relies heavily on the identification
of functions and objects introducing entropy into the script context. Values
classified as entropy are overwritten in the \twin with the value received
from the \orig. This ensures that functions such as \texttt{Math.random} and
\texttt{Date.now} return the exact same value, which is crucial for synchronous
execution. While it is obvious for \texttt{Date.now}, it is not immediately
clear for other methods. Therefore, \emph{entropy inducing} methods are detected and
filtered incremetally during runtime. Hence, if a detection has triggered but
the cause was not an \infoleak{}, it is included into the list of entropy
methods.
\paragraph{Rendezvous and Checking Points.}
Vital program points where \orig{} and \twin{} are synchronized are bytecode
handler functions. If a handler function returns data into the script context, it
is first determined if the handler function is an entropy inducing
function. However, the vast majority of function invocations and object accesses
do not introduce entropy and are checked for equality between \orig and \twin
on the fly. If a difference is encountered that is not classified
as entropy, we assume that an information leak occurred and take actions, namely
logging the incident and terminating both processes.

\subsection{Chakra Instrumentation}
\label{implementation:sync}

The Chakra JavaScript Engine contains a JIT compiler. It runs in a dedicated thread, identifies frequently executed (so called \textit{hot}) functions and compiles them to native code.
Our current implementation works on script interpreters, hence we disabled the
JIT compiler. This is currently a prototype limitation whose solution we
discuss in Section~\ref{discussion:limit}.

In order to synchronize execution and check for information leaks, we instrumented the main loop of the Chakra interpreter, which is located in the \texttt{Js::Interpreter\-Stack\-Frame::Process} function. It is invoked recursively for each JavaScript call and iterates over the variable length bytecodes of the JavaScript function. The main loop contains a \texttt{switch} statement, which selects the corresponding handler for the currently interpreted bytecode. The handler then operates on the JavaScript context dependent on the operands and the current state. In the examined Chakra versions, we observed up to 648 unique bytecodes. Prior to the invocation of a bytecode handler, our instrumentation transfers the control flow to a small, highly optimized assembly stub, which decides whether the current bytecode is vital for our framework to handle.

We intercept all \texttt{call} and \texttt{return} as well as necessary \texttt{conversion} bytecodes in order to extract metadata such as JavaScript function arguments, return values, and conversion values. \texttt{Conversion} bytecodes handle dynamic type casting, native value to JavaScript object and JavaScript object to native value conversions. Additionally, we intercept engine functions that handle implicit type casts at native level, because they are invoked by other bytecode handlers as required and have no bytecode equivalents themselves.
Furthermore, all interception sites support the manipulation of the outgoing native value or JavaScript object for the purpose of entropy elimination in the JavaScript context of the \twin process.

\subsection{AVM Instrumentation}
\label{implementation:flash}

Instrumentation of the AVM is based on prior work of F-Secure~\cite{avmfunprofit2014}
and Microsoft~\cite{insideavm2012}. We hook at the end of the native method
\texttt{verifyOnCall} inside \texttt{verifyEnterGPR} to intercept ActionScript method calls and retrieve ActionScript method names. At these points,
\orig and \twin can be synchronized. Parameters flowing into an ActionScript
method and return data flowing back into the ActionScript context can be
dissected, too. They are also processed inside the method \texttt{verifyEnterGPR}.
Based on their high level ActionScript types, the parameters and return
data can be compared in the \orig and \twin.
This way, we can keep the \orig and \twin in sync at method calls, check for
information leaks and mediate entropy data from the \orig to the \twin.

\section{Evaluation}
\label{eval}

In the following, we present evaluation results for our prototype implementation of \toolname{} in the form of performance and memory usage benchmarks.
The benchmarks were conducted on a system running Windows 8.0/8.1 that was
equipped with a 4th generation Intel i7-4710MQ quad-core CPU and 8GB DDR3
RAM. Furthermore, we demonstrate how our prototype can successfully detect
several kinds of real-world \infoleak{}s. 

\subsection{Re-randomization of Process Modules}
\label{eval:rerand}
We evaluated our re-randomization engine according to its effectiveness, memory
usage, and performance.
\paragraph{Effectiveness.}

\begin{figure*}[tb]
	\centering
	\includegraphics[width=0.90\linewidth]{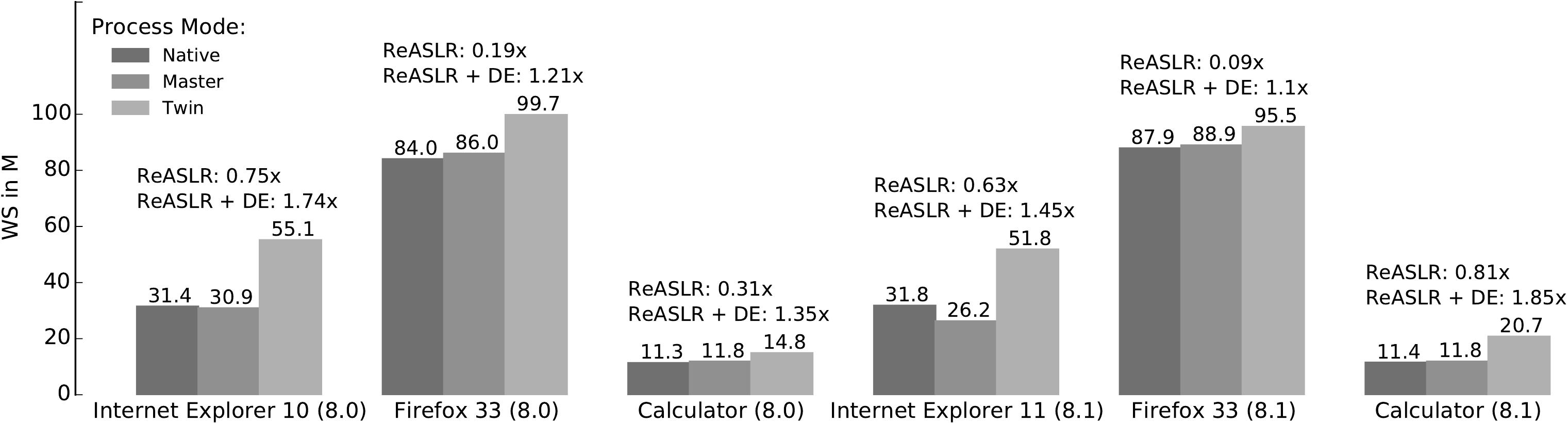}
    \caption{Memory overhead of re-randomization and dual execution measured via
    working set (WS) consumption in megabytes (M): Native processes on Windows 8.0 and 8.1 are contrasted to their
    counterparts running in re-randomized dual execution mode (master and twin).}
    \label{fig:ws_overhead}
\end{figure*}

We applied re-randomization to internal Windows
applications and third-party applications, to verify that modules in the \twin
are based at different addresses than in the \orig.  We therefore compared base
addresses of all loaded images between the two processes and confirmed that
all images in the \twin process had a different base address than in the \orig, except
\texttt{ntdll.dll}.  See the discussion in Section \ref{discussion:limit} for details on
the difficulties of remapping the 64-bit and 32-bit NT Layer DLLs.
The extended version of this paper lists important Windows DLLs, re-randomized in different processes 
running \emph{simultaneously} on a \emph{single} user
session~\cite{TR_INFO_LEAKS}.
\paragraph{Physical Memory Usage.}
To inspect the memory overhead of our re-ran\-dom\-iza\-tion scheme, we measured the
working set characteristics for different \orig and re-randomized \twin
processes compared to native processes.
Figure~\ref{fig:ws_overhead} shows the memory working sets of three
applications.
\emph{ReASLR} denotes thereby the re-randomization within a single process.
\textsc{DE} means that two processes are running, whereby the
\orig{}'s randomization is kept native while the \twin{} is re-randomized. The applications besides IE are only included to measure the memory overhead and are not synchronized.
We calculate the memory overhead of \raslr{} (\emph{ReASLR}) of a \emph{single} process as follows:

\[ Overhead(ReASLR) = \frac{WS(Twin)}{WS(Native)}-1 \]

Thus, the overall memory overhead based on working sets is 0.46 times.
When running a program or process in \raslr \emph{and} \dexec (\emph{DE}), we have to include 
both \orig and \twin into the memory overhead calculation. Therefore, the overhead is
calculated by

\[ Overhead(ReASLR + DE) = \frac{WS(Twin) + WS(Master)}{WS(Native)}-1 \]
 
Its overall value is 1.45 times.
Note that memory working sets can highly vary during an application's runtime,
and thus, are difficult to quantify. The measurements shown in
Figure~\ref{fig:ws_overhead}
were performed after the application has finished startup, and was waiting
for user input (i.e., it was idle and all modules were loaded and initialized).
Due to additional \twin{}s for \orig processes, the overall additional memory is
about one to two times per protected process.
The technical report provides more details on the working set
characteristics~\cite{TR_INFO_LEAKS}.
\paragraph{Re-randomization and Startup Time.}
When a program is started the first time after a reboot, the kernel needs to
create section objects for image modules. Hence, the first start of a program
always takes longer than subsequent starts of the same program.  To measure the
additional startup and module load times our protection introduces, we first
run each program natively once to allow the kernel to create section objects of
most natively used DLLs, and close it afterwards. We then start the program
natively without protection and measure the time until it is idle and all of
its initial modules are loaded.
In the same way, we measure the time from process creation until both the
\orig and \twin process have their inital modules loaded.
\begin{table*}[b!t!]
    \centering
    \caption{Startup times in seconds and startup slowdowns of native 32-bit
        applications compared to their counterparts running with \raslr and
        \dexec on Windows 8.0 and Windows 8.1 (both 64-bit).
    }
    ~\\
        \ssmall
    \tabcolsep=0.09cm
    \begin{tabular}{ l| c c c| c c c c}
            & Native (8.0) & ReASLR+DE (8.0) & Slowdown & Native (8.1) & ReASLR+DE (8.1) & Slowdown \\ 
        \midrule                                                                         
        IE tab spawn  & 0.9163 s & 2.0710 s & 1.3x  & 0.5194 s & 1.3082 s & 1.5x  \\
        Firefox       & 0.9624 s & 1.8064 s & 0.9x  & 1.3823 s & 1.5441 s & 0.1x  \\
        Calculator    & 0.3484 s & 0.3610 s & 0.0x  & 0.4391 s & 0.6599 s & 0.5x  \\
        \bottomrule
    \end{tabular}
    \label{tab:goodtimes_loadtimes}
\end{table*}
The startup comparison can be seen in Table~\ref{tab:goodtimes_loadtimes}.
As expected, the startup times of applications protected with our approach are
approximately doubled. This is caused by the fact that a \twin process needs to be spawned for each \orig that should
be protected.

\subsection{Detection Engine}
\label{sec:eval:de}
Next, we evaluate the impact of \toolname on the user experience and its
effectiveness in detecting \infoleak{}s.
\begin{table*}[b!t!]
    \centering
    \caption{
        Native script execution of IE 11 on Windows 8.1 64-bit
        compared to the script execution of IE 11 instrumented with
        \toolname. Execution time is measured in milliseconds using the internal F12
        developer tools provided by IE.
    }
    ~\\
        \ssmall
        \tabcolsep=0.07cm
    \begin{tabular}{ l |l l l l l l l l l l l l l l l l l l l l}
         Website & \rotatebox{79}{google.com} & \rotatebox{79}{facebook.com} & \rotatebox{79}{youtube.com} & \rotatebox{79}{yahoo.com} & \rotatebox{79}{baidu.com} & \rotatebox{79}{wikipedia.org} & \rotatebox{79}{twitter.com} & \rotatebox{79}{qq.com} & \rotatebox{79}{taobao.com} & \rotatebox{79}{linkedin.com} & \rotatebox{79}{amazon.com} & \rotatebox{79}{live.com} & \rotatebox{79}{google.co.in} & \rotatebox{79}{sina.com.cn} & \rotatebox{79}{hao123.com} \\
        \midrule
        Native & 425 & 774 & 1196 & 3674 & 1108 & 472 & 599 & 2405 & 645 & 439 & 958 & 254 & 483 & 3360 & 373 \\
        \toolname{} & 482 & 961 & 1519 & 4722 & 1339 & 513 & 623 & 2724 & 824 & 517 & 1210 & 275 & 517 & 4269 & 379 \\
        Overhead & 13.4\% & 24.1\% & 27\% & 28.5\% & 20.8\% & 8.6\% & 4\% & 13.2\% & 27.7\% & 17.7\% & 26.3\% & 8.2\% & 7\% & 27\% & 1.6\% \\
        \bottomrule

    \end{tabular}
    \label{tab:script_exec}
\end{table*}

\paragraph{Script Execution Time and Responsiveness.}
We used the 15 most visited websites worldwide~\cite{alexaTop} to test how the
current prototype interferes with the normal usage of these pages.
Besides the subjective impression while using the page, we utilized the F12
developer tools of Internet Explorer 11 to measure scripting execution time
provided by the \emph{UI Responsiveness} profiler tab. These tests were
performed using Windows 8.1 64-bit and Internet Explorer 11.
While we introduce a
performance hit of around 17.0\% on average, the subjective user experience was
not noticeably affected.
This is due to IE's deferred parsing, which results in displaying content to the user before all computations have finished.

\paragraph{Information Leak Detection.}

We tested our approach on a pure memory disclosure vulnerability (CVE-2014-6355)
which allows illegitimately reading data due to a JPEG parsing flaw in Microsoft's Windows
graphics component~\cite{zalewski2014_CVE_2014_6355}. It can be used to defeat ASLR by
reading leaked stack information back to the attacker via the \texttt{toDataURL} method of a
\texttt{canvas} object. We successfully detected this leak at the point of the call to
\texttt{toDataURL} in the \orig and \twin process. In the same way, detection was
successful for an exploit for a similar bug
(CVE-2015-0061~\cite{zalewski2015_CVE_2015_0061}).

To further verify our prototype, we evaluated it against an exploit for CVE-2011-1346, 
a vulnerability that was used in the pwn2own contest 2011 to bypass
ASLR~\cite{thezdi2011_CVE_2011_1346}. As this memory disclosure bug is specific for IE~8,
we ported the vulnerability into IE 11. An uninitialized
\texttt{index}
attribute of a new HTML \texttt{option} element is used to leak information.
Similarly, we successfully detected this exploitation attempt when the \texttt{index} attribute was accessed.

Additionally, we tested our prototype on another real-world vulnerability (CVE-2014-0322) that was used
in targeted attacks~\cite{cve_2014_0322}.
It is a use-after-free error that can be utilized to
increase an arbitrary bit, which is enough to create \infoleak{}s.
\toolname triggered as a Vtable pointer was returned into the JavaScript context.
Therefore, the \infoleak was detected successfully.

We also constructed a toy example in which our native code creates an
\infoleak by overwriting the length field of an array. Additionally, the image
base of \texttt{jscript9.dll} is written after the array data. In our tests, we reliably detected the out-of-bounds
read of the image base and stopped the execution of the process. Exploit
details are provided in the technical report~\cite{TR_INFO_LEAKS}.

\paragraph{False Positive Analysis.}
We analyzed the 100 top websites
worldwide~\cite{alexaTop} to evaluate if our prototype can precisely handle
real-world, complex websites and their JavaScript contexts without triggering
false alarms.  None of the tested websites did generate an alert, indicating
that the prototype can accurately synchronize the master and twin process.

\section{Related Work} \label{related}
In the following, we review work closely related to ours and discuss differences
to our approach.
\paragraph{Randomization Techniques.}
Several approaches have been proposed to either improve address space layout
randomization, randomize the data space, or randomize on single instruction
level. For example, binary stirring~\cite{wartell2012binary} re-randomizes code pages at a high rate for a high
performance cost. While it hinders attackers to \emph{use} \infoleak{}s in
code-reuse attacks, it does not impede their creation by itself. In
contrast, our re-randomization scheme reuses the native operating system loader
and is the base to allow \infoleak{} detection with \dexec. 
Other solutions~\cite{pappas2012smashing, kil2006address, pappas2012smashing} are prone to
JIT-ROP code-reuse attacks~\cite{snow2013just}, which are based on \infoleak{}s. Address space layout
permutation is an approach to scramble all data and functions of a
binary~\cite{kil2006address}. Therefore, a given ELF binary has to be rewritten
and randomization can be applied on each run. ORP~\cite{pappas2012smashing} rewrites instructions of a given
binary and reorders basic blocks. As discussed above, it is prone to \infoleak
attacks, which we detect. Instruction set randomization~\cite{barrantes2003randomized, kc2003countering}
complicates code-reuse attacks as it encrypts code pages and
decrypts it on the fly. However, in the presence of \infoleak{}s combined with
key guessing~\cite{snow2013just, sovarel2005s, weiss2006known} it can be circumvented.  Instruction layout
randomization (ILR)~\cite{hiser2012ilr} randomizes the location of each instruction on
each run, but no re-randomization occurs. Thus, the layout can be reconstructed
with the help of an \infoleak.
\emph{Readactor} is a defensive system that aims to be resilient against
just-in-time code-reuse attacks~\cite{crane2015readactor}. It hides code pointers behind execute-only
trampolines and code itself is made execute-only, to prevent an attacker building
a code-reuse payload just-in-time. 
However, it has been shown that it is vulnerable against an attack named \emph{COOP}, which reuses virtual functions~\cite{schuster2015counterfeit}. Unlike Readactor, \toolname prevents COOP, as this attack needs an \infoleak{} as first step. Crane et al. recently presented an enhanced version of Readactor, dubbed \emph{Readactor++}~\cite{crane2015readactorplusplus}, that also protects against whole function reuse attacks such as COOP. This is achieved through function pointer table randomization and insertion of booby traps. Consequently, an adversary can no longer obtain meaningful code locations that can be leveraged for code-reuse attacks. Readactor++ also does not detect or prevent the exploitation of memory disclosures, which poses a potential attack vector.
\paragraph{Multi-Execution Approaches.}
\label{_sec:related_n_variant}

Most closely related to our research are \emph{n-variant systems}, which run
variants of the same program with diverse memory layout and
instructions~\cite{Cox:2006:NSS}.
Similar work
runs \emph{program replic{\ae}} synchronized at system calls to demonstrate the
detection of memory exploits against the lightweight server \texttt{thttpd} on
the Linux platform~\cite{bruschi2007diversified, varan:asplos15}.

The major drawback of theses systems is the detection approach: if a
memory error is abused, one of the variants eventually crashes, which indicates
an attack. As \infoleak{}s \emph{do not} constitute a memory error, they
\emph{do not} raise any exception-based signal. Thus, they remain undetected in
these systems.
One significant implication is that
unlike \toolname, n-variant systems do not protect against just-in-time
code-reuse attacks such as JIT-ROP~\cite{snow2013just}. Similarily, this is the
case with COOP attacks in browsers~\cite{schuster2015counterfeit}.
N-variant systems
prevent conventional ROP attacks~\cite{Prandini2012return, volckaert2015cloning} with multi process
execution and disjunct virtual address spaces:
An attacker supplied absolute
address (e.g., obtained through a remote memory disclosure vulnerability) is
guaranteed to be invalid in $n-1$ replicas.
Hence, any system call utilizing this address will trigger a detection. However,
JIT-ROP attacks may performs several memory disclosures and malicious
computations without executing a system call inbetween, and thus, can evade
traditional n-variant systems. COOP attacks may as well perform touring-complete
computations on disclosed memory without executing a system call and evade these systems.

\section{Discussion}
\label{discussion}

In the following, we discuss potential shortcomings of our approach and the prototype, and also sketch how these
shortcomings can be addressed in the future.
\paragraph{Further Information Leaks.}
Serna provided an in-depth overview of techniques that utilize {\infoleak}s for exploit development~\cite{serna2012info}.
The techniques he discussed during the presentation utilize JavaScript code.
As our prototype leverages the JavaScript engine of the browser itself, each information leak that is based on these techniques is detected.
This implies
that memory disclosure attacks that leverage other (scripting) contexts (e.g., VBScript) can potentially
bypass our implementation.
However, in practice exploits are typically triggered via JavaScript and thus
our prototype can detect such attacks.
Furthermore, due to the generic nature of our approach, our current prototype can
be extended by instrumenting other scripting engines as well.
\paragraph{Prototype Limitations.}
\label{discussion:limit}
In the unlikely event one of the functions we classified as entropy source, such as \texttt{Math.random} or \texttt{Date.now}, contain a memory disclosure bug, our approach can lead to an under-approximation of detected information leaks. In this specific case, the \orig confuses the leaked pointer with data from the entropy source and transfers it to the \twin process. This is an undesirable state, because \toolname does not prevent the memory layout information to leak into the script context. However, the obtained pointer is only valid in the \orig process. An attempt to leverage the pointer to mount a code-reuse attack crashes the \twin. As a consequence, \toolname halts the \orig process and prevents further damage.

The current prototype disables the JIT Engine as we protect the interpreter
only. However, \emph{dynamic binary instrumentation} (DBI~\cite{luk2005pin, bruening2001design}) frameworks allow to
synchronize processes on the intruction or basic block level, and hence, make it
possible to hook emitted JIT code to dispatch our assembly stub in order to
synchronize and check within the JIT code.

Asynchronious JavaScript events are currently not synchronized. This is solvable with DBI
frameworks as well: If an event triggers in the \orig{} process, we let the \twin{} execute to the same point.
Then \toolname{} sets up and triggers the same event in the \twin{} process.

One additional shortcoming of our prototype implementation is the identical mapping of
\texttt{ntdll.dll} in all processes. As this DLL is initialized already at startup,
remapping it is a cumbersome operation. 
JavaScript, HTML, and other contexts in browsers normally do not interact
directly with native \texttt{ntdll.dll} Windows structures,
and thus internal JavaScript objects, do not contain direct memory references to it.
Hence, attackers resort to disclose addresses from libraries other than \texttt{ntdll.dll} at first.
On the contrary, there might be script engines which directly interact with \texttt{ntdll.dll}.  
Still, the issue is probably solvable with a driver loaded during boot time.

Another technical drawback is the application of re-randomization on every
process on the OS, as DLL modules of each process would turn into non-shareable
memory and increase physical memory consumption. This can be avoided by
protecting only critical processes that represent a valid target for attacks.

\paragraph{Deployment.}

The current prototype is not meant to be a protection framework for end users of web browsers.
It is intended to be deployed as a system for scanning web pages
to discover unknown exploits which utilize information leaks. As ASLR needs to be circumvented as a 
first step of each modern exploit against web browsers, \toolname{} has the
advantage to provide an early detection of the exploit process.

\section{Conclusion}
\label{conclusion}

Over the last years, script engines were used to exploit vulnerable applications.
Especially web browsers became an attractive target for a plethora of
attacks. State-of-the-art vulnerability exploits, both
in academic research~\cite{goktas2014out, davi2014stitching, goktacs2014size,
carlini2014rop, schuster2014evaluating} and in-the-wild~\cite{yu2014write,
yu2014rops, yan2014art, vupenFox, browserweakestbyte}, rely on memory disclosure attacks.

In this work, we proposed a fine-grained, automated scheme to reliably detect such \infoleak{}s in script engines.
It is based on the insight that \infoleak{}s result in a noticable difference in
the script context of two synchronized processes with different randomization.
We implemented a prototype of this idea for the proprietary browser 
IE to demonstrate that our approach is viable even on closed-source
systems. An empirical evaluation demonstrates that we can reliably detect
real-world attack vectors and that the approach induces a moderate performance
overhead only (around 17\% overhead on average).
While most research focused on mitigating specific types of vulnerabilities, we
address the root cause behind modern attacks since most of them rely on \infoleak{}s as a first step. Our approach thus serves as another defense layer to complement defenses such as DEP and ASLR.

\subsection*{Acknowledgements}
We would like to thank the anonymous reviewers for their valuable comments. 
This work was supported by the European Commission through the ERC Starting
Grant No. 640110 (BASTION).

\bibliographystyle{abbrv}
\bibliography{bibliography}

\end{document}